\documentclass[aps,prc,twocolumn,showpacs,preprintnumbers,amsmath,amssymb,floatfix,superscriptaddress]{revtex4}

\usepackage{graphicx}% Include figure files
\usepackage{dcolumn}% Align table columns on decimal point
\usepackage{bm}% bold math
\usepackage{booktabs}
\usepackage{longtable}

\begin{document}

%\preprint{APS/123-QED}

\title{Cross Sections of the $^{83}$Rb(p,$\gamma$)$^{84}$Sr and $^{84}$Kr(p,$\gamma$)$^{85}$Rb Reactions at Energies Characteristic of the Astrophysical $\gamma$ Process}% Force line breaks with \\
%\thanks{A footnote to the article title}%

\author{M. Williams}
\affiliation{TRIUMF, Vancouver, British Columbia, V6T 2A3, Canada}
\affiliation{Department of Physics, University of York, Heslington, York, YO10 5DD, United Kingdom}

\author{B. Davids}
\affiliation{TRIUMF, Vancouver, British Columbia, V6T 2A3, Canada}
\affiliation{Department of Physics, Simon Fraser University, Burnaby, British Columbia V5A 1S6, Canada}

\author{G. Lotay}
\affiliation{Department of Physics, University of Surrey, Guildford, GU2 7XH, United Kingdom}

\author{N. Nishimura}
\affiliation{Astrophysical Big Bang Laboratory, CPR, RIKEN, Wako, Saitama 351-0198, Japan}
\affiliation{Nishina Center for Accelerator-Based Science, Wako, Saitama 351-0198, Japan}

\author{T. Rauscher}
\thanks{ORCID: 0000-0002-1266-0642}
\affiliation{Department of Physics, University of Basel, Klingelbergstr.\ 82, CH-4056 Basel, Switzerland}
\affiliation{Centre for Astrophysics Research, University of Hertfordshire, Hatfield AL10 9AB, United Kingdom}

\author{S.\ A.\ Gillespie}
\thanks{Present Address: FRIB, Michigan State University, East Lansing, MI, 48824, USA}
\affiliation{TRIUMF, Vancouver, British Columbia, V6T 2A3, Canada}

\author{M. Alcorta}
\affiliation{TRIUMF, Vancouver, British Columbia, V6T 2A3, Canada}

\author{A.\ M.\ Amthor}
\affiliation{Department of Physics and Astronomy, Bucknell University, Lewisburg, PA, 17837, USA}

\author{G.\ C. Ball}
\affiliation{TRIUMF, Vancouver, British Columbia, V6T 2A3, Canada}

\author{S.\ S.\ Bhattacharjee}
\affiliation{TRIUMF, Vancouver, British Columbia, V6T 2A3, Canada}

\author{V. Bildstein}
\affiliation{Department of Physics, University of Guelph, Guelph, Ontario, N1G 2W1, Canada}

\author{W.\ N. Catford}
\affiliation{Department of Physics, University of Surrey, Guildford, GU2 7XH, United Kingdom}

\author{D.\ T.\ Doherty}
\affiliation{Department of Physics, University of Surrey, Guildford, GU2 7XH, United Kingdom}

\author{N.\ E.\ Esker}
\altaffiliation{Present Address: Department of Chemistry, San Jose State University, San Jose, CA 95192, USA}
\affiliation{TRIUMF, Vancouver, British Columbia, V6T 2A3, Canada}

\author{A.\ B.\ Garnsworthy}
\affiliation{TRIUMF, Vancouver, British Columbia, V6T 2A3, Canada}

\author{G. Hackman}
\affiliation{TRIUMF, Vancouver, British Columbia, V6T 2A3, Canada}

\author{K. Hudson}
\affiliation{TRIUMF, Vancouver, British Columbia, V6T 2A3, Canada}
\affiliation{Department of Physics, Simon Fraser University, Burnaby, British Columbia V5A 1S6, Canada}

\author{A. Lennarz}
\affiliation{TRIUMF, Vancouver, British Columbia, V6T 2A3, Canada}

\author{C. Natzke}
\affiliation{TRIUMF, Vancouver, British Columbia, V6T 2A3, Canada}
\affiliation{Department of Physics, Colorado School of Mines, Golden, CO 80401, USA}

\author{B. Olaizola}
\altaffiliation{Present Address: CERN, CH-1211 Geneva 23, Switzerland}
\affiliation{TRIUMF, Vancouver, British Columbia, V6T 2A3, Canada}

\author{A. Psaltis}
\altaffiliation{Present Address: Institut f{\"u}r Kernphysik, Technische Universit{\"a}t Darmstadt, Darmstadt,  D-64289, Germany}
\affiliation{Department of Physics and Astronomy, McMaster University, Hamilton, Ontario, L8S 4L8, Canada}

\author{C.\ E.\ Svensson}
\affiliation{Department of Physics, University of Guelph, Guelph, Ontario, N1G 2W1, Canada}

\author{J. Williams}
\affiliation{TRIUMF, Vancouver, British Columbia, V6T 2A3, Canada}

\author{D. Walter}
\affiliation{TRIUMF, Vancouver, British Columbia, V6T 2A3, Canada}
\affiliation{Department of Astronomy and Physics, Saint Mary's University, Halifax, Nova Scotia, B3H 3C3, Canada}

\author{D. Yates}
\affiliation{TRIUMF, Vancouver, British Columbia, V6T 2A3, Canada}
\affiliation{Department of Physics and Astronomy, University of British Columbia, Vancouver, BC V6T 1Z4, Canada}

\date{\today}% It is always \today, today,
             %  but any date may be explicitly specified

\begin{abstract}
We have measured the cross section of the $^{83}$Rb(p,$\gamma)^{84}$Sr radiative capture reaction in inverse
kinematics using a radioactive beam of $^{83}$Rb at incident energies of 2.4 and $2.7 A$ MeV. Prior to the radioactive beam measurement, the $^{84}$Kr(p,$\gamma)^{85}$Rb radiative capture reaction was measured in inverse
kinematics using a stable beam of $^{84}$Kr at an incident energy of $2.7 A$ MeV.
The effective relative kinetic energies of these measurements lie within the relevant energy window for the $\gamma$ process in supernovae.
The central values of the measured partial cross sections of both reactions were found to be $0.17-0.42$ times the predictions of statistical model calculations. Assuming the predicted cross section at other energies is reduced by the same factor leads to a slightly higher calculated abundance of the $p$ nucleus $^{84}$Sr, caused by the reduced rate of the $^{84}$Sr($\gamma$,p)$^{83}$Rb reaction derived from the present measurement.
\end{abstract}

% PACS, the Physics and Astronomy
                             % Classification Scheme.
%\keywords{Suggested keywords}%Use showkeys class option if keyword
                              %display desired
\maketitle

%\tableofcontents

\section{\label{sec:intro}Introduction}

More than 60 years have elapsed since it was established that the stellar nucleosynthesis of elements heavier than iron is largely governed by the ($s$)low and ($r$)apid neutron capture processes \cite{B2FH,cameron57}. However, there are some 30 stable, neutron-deficient nuclides between Se and Hg that cannot be formed by either of these processes and their astrophysical origin remains a subject of active investigation \cite{Rauscher,nobsnIa}. As these $p$ nuclides only account for a small fraction of overall elemental abundances, they are not directly observable in stars or supernova remnants. Hence, it is necessary to study their formation using a combination of detailed nucleosynthetic models and meteoritic data \cite{Rauscher2}.

Presently, $p$ nuclides are thought to be formed by photodisintegration reactions on pre-existing $r$- and $s$-process
seed nuclei in the O/Ne layers of core-collapse supernovae (ccSNe) \cite{Arnould,howard} and in thermonuclear supernovae \cite{trav11,nobsnIa}, with typical peak plasma temperatures of $T_{max}\sim2-3.5$ GK in the $p$-process layers. In particular, ($\gamma,n$) reactions drive the pathway of nucleosynthesis toward the neutron-deficient side of stability until neutron separation energies become high enough that ($\gamma,p$) and ($\gamma,\alpha$) reactions largely dominate the flow of material. This astrophysical $\gamma$ process is capable of reproducing the bulk of the $p$ nuclides within a single stellar site \cite{Rauscher2}. However, there are abiding issues in obtaining abundances consistent with solar system values for the lightest $p$ nuclides having mass number $A$ $\lesssim$ 110 \cite{arngor,umberto} that have yet to be resolved. These discrepancies may be addressed through changes to the underlying nuclear physics input, as cross sections of $\gamma$-process reactions are almost entirely unmeasured and the related reaction rates are based exclusively on theoretical calculations.  

It is well known (see, e.g., Ref.\ \cite{holmes}) that for most reactions on intermediate and heavy targets the impact of thermal excitations of target nuclei in the stellar plasma is smaller in the direction with a positive reaction $Q$ value than in the inverse, endothermic direction. This means that reactions on the ground state of a target nucleus make a larger relative contribution to the total astrophysical reaction rate in the exothermic direction than do inverse reactions on the ground state of the product nucleus to the total astrophysical reaction rate in the endothermic direction. Notable exceptions to this so-called ``$Q$-value rule" for astrophysical reaction rates are capture reactions, for which the relative contributions of thermally excited states are always smaller in the capture direction of the reaction than in the photodisintegration direction, regardless of the $Q$ value \cite{qval,Rauscherb}. For the nucleosynthesis of $p$ nuclides, this implies that it is more advantageous to experimentally study radiative capture reactions rather than the inverse photodisintegration reactions,
whenever a direct constraint of the reaction rate is attempted \cite{Rauscher2}. The vast majority of these reactions involve unstable nuclei and exhibit cross sections of order 100 $\mu$b at the most important energies. As such, most $\gamma$-process reactions remain experimentally inaccessible, notwithstanding the latest developments in the production and acceleration of radioactive ion beams. Hence, astrophysical abundance calculations have relied extensively on the use of the Hauser-Feshbach (HF) theory of the statistical model \cite{NONSMOKER1,NONSMOKER2}. Although this approach is valid for reactions important for the synthesis of $p$ nuclides, the nuclear properties required as input are not well known for nuclei outside the valley of $\beta$ stability. This lack of information  leads to uncertainties in the predictions of astrophysical reaction rates. Therefore experimental cross section measurements are required. Here, we describe a direct measurement of the cross section of a $\gamma$-process reaction involving an unstable nuclide in the relevant energy window for the $\gamma$ process, which covers relative kinetic energies $E_{cm}$ from approximately $1.4-3.3$~MeV 
\cite{Rauscher3,nobsnIa}. 

The measurement performed at the ISAC-II facility of TRIUMF first reported in Ref.\ \cite{lotay21} utilized an intense, radioactive beam of $^{83}$Rb ions, together with the TIGRESS $\gamma$-ray detector array \cite{TIGRESS} and the EMMA recoil mass spectrometer \cite{EMMA}, to investigate the cross section of the $^{83}$Rb($p$,$\gamma$)$^{84}$Sr reaction. By exploiting the fact that the electromagnetic decay of proton-unbound states in $^{84}$Sr, populated via resonant proton capture on the 5/2$^{-}$ ground state of $^{83}$Rb, predominantly proceeds via $\gamma$-decay cascades to the lowest-lying 2$^{+}$ level rather than directly to the ground state, we inferred the total reaction cross section from the observed 793.22(6)-keV, 2$^{+}_1$ $\rightarrow$ 0$^{+}_1$ $\gamma$-ray yield \cite{Singh}. It was suggested that the $^{83}$Rb($p$,$\gamma$)$^{84}$Sr reaction rate has a substantial influence on the calculated $^{84}$Sr abundance obtained in ccSNe \cite{Rapp,Rauscher}. Recently, elevated levels of $^{84}$Sr have been discovered in calcium-aluminium-rich inclusions (CAIs) in the Allende meteorite \cite{Charlier}. While this may be accounted for by $r$- and $s$-process variability in $^{88}$Sr production, another possible resolution might be increased production of $^{84}$Sr in the astrophysical $\gamma$ process.

\section{\label{sec:exp}Experimental Procedures}

To produce the radioactive ion beam, we bombarded a ZrC production target with 500 MeV protons from the TRIUMF cyclotron at currents of up to 50 $\mu$A. In the experiment, surface-ionized $^{83}$Rb ions with a half life $T_{1/2}=86.2(1)$~d \cite{mccutchan} were accelerated and stripped to the 23$^+$ charge state before reaching energies of $2.4 A$ and $2.7 A$~ MeV in the ISAC-II facility \cite{laxdal14}. They were directed onto 300 to 900~$\mu$g~cm$^{-2}$ thick polyethylene (CH$_2$)$_n$ targets at intensities of $1-5\times10^7$~s$^{-1}$ to measure the $p(^{83}$Rb,$\gamma)^{84}$Sr reaction cross section. The beam intensity was limited by the power that could be dissipated by the reaction target via thermal radiation. Prior to the radioactive beam study, a measurement of the $p(^{84}$Kr,$\gamma)^{85}$Rb radiative capture cross section was carried out at a bombarding energy of $2.7A$~MeV and similar intensities. This was used as a test of the new experimental setup with a stable beam of comparable mass free from radioactive-beam-induced background. Measurements with the Faraday cup at the EMMA target position showed that the beam spots were stable over time and were fully contained within a circular aperture of 1~mm radius centred on the beam axis.

Prompt $\gamma$ rays were detected with 12 Compton-suppressed HPGe detectors of the TIGRESS array, while the radiative capture products $^{85}$Rb and $^{84}$Sr were transmitted to the focal plane of the EMMA recoil mass spectrometer in either the $25^+$ or $26^+$ charge state. Eight of the HPGe detectors were centred at 90$^\circ$ and four were placed at 135$^\circ$ with respect to the beam direction. All were positioned 11~cm from the target. Electrostatic potential differences of 320~kV across the gaps of the two electrostatic deflectors were maintained. An electromagnetic separator capable of transporting ions with an electrostatic rigidity of 13 MV was needed to transmit the recoils of these reactions. The rigidity limits of EMMA make the spectrometer well matched to recoil energies typical in $\gamma$-process studies.

The recoils of these radiative capture reactions were strongly forward focussed, with a maximum recoil angle of 0.1$^\circ$ due to the inverse kinematics. Multiple scattering in the target foils broadened the distributions with a planar scattering angle characterized by a Gaussian with a standard deviation of approximately 0.25$^\circ$. A rectangular aluminum entrance aperture 8~cm downstream of the target limited the horizontal and vertical projections of the recoil scattering angle to $\pm1.2^\circ \times \pm1.2^\circ$ in order to reduce the number of elastically scattered beam ions transmitted through the spectrometer. Two slit systems symmetrically located upstream and downstream of the dipole magnet of EMMA were narrowed to a width of $\pm3$~cm to limit the energy acceptance of the spectrometer, and the final slit system at the mass/charge (m/q) dispersed focal plane was opened to a width of 6~mm, corresponding to a m/q acceptance of $\pm0.3\%$. Together, the slit systems and other components of the spectrometer reduced the rate of scattered beam reaching the recoil detectors by a factor of 50~000. Recoils and scattered beam ions passing through the focal plane slit system of the spectrometer traversed a parallel grid avalanche counter and a transmission ionization chamber before stopping in a 3000~mm$^2$, 500~$\mu$m thick ion-implanted Si detector.

\subsection{\label{sec:beamnorm}Luminosity Determination}

During the radioactive beam experiment, the $^{83}$Rb beam was accompanied by a significant $^{83}$Sr component. Typically, the composition of the beam would be determined by energy loss measurements using a Bragg ionization detector, as described in Ref.\ \cite{cruz19}. However, at the low bombarding energies of the present study, this method could not distinguish $^{83}$Rb and $^{83}$Sr ions. The beam composition was instead determined by $\gamma$-ray spectroscopic analysis using the decays of elastically scattered beam ions that stopped in the removable entrance aperture of EMMA throughout the experiment. Immediately following the measurement, the aperture was removed and installed within the GRIFFIN spectrometer \cite{GRIFFIN}, which was used to measure $\gamma$ rays emitted following the $\beta$ decays of both $^{83}$Rb and $^{83}$Sr, which has a $T_{1/2} = 32.41(3)$~h \cite{mccutchan}. A second measurement was performed 22 days following the experiment. On the basis of these measurements, the radioactive ion beam was found to be $62(3)\%$ $^{83}$Rb. Throughout the cross section measurement, the 762.65(10) keV transition from the 804.77(3) keV state to the 42.078(2) keV state in $^{83}$Rb that follows the EC/$\beta^+$ decay of $^{83}$Sr was continuously observed using the TIGRESS array, allowing us to determine its energy resolution to be 2.5~keV~(FWHM) at 763~keV.

%Spectra obtained from both the initial and second measurements are displayed on Figure ..., with the $\gamma$-rays of interest indicated. 

% Insert sentences about analysis procedure. How was the ROI for each peak defined? Which gamma-rays were used? How many counts were detected in each measurement? Perhaps exponential decay equation

% What is the breakdown of systematic vs statistical uncertainty in the above measurement? What are the dominant sources of systematic error.

%Other potential sources of systematic error were considered, such as difference in the mean scattering angle and implantation depth of $^{83}$Rb vs $^{83}$Sr (both arising due to differing atomic number, $Z$). 

%Both being alkaline metals, we assumed that any potential for migration of the implanted ions within the aluminium would be negligible and/or equivalent.  

\begin{figure}[t!]
    \includegraphics[width=\linewidth]{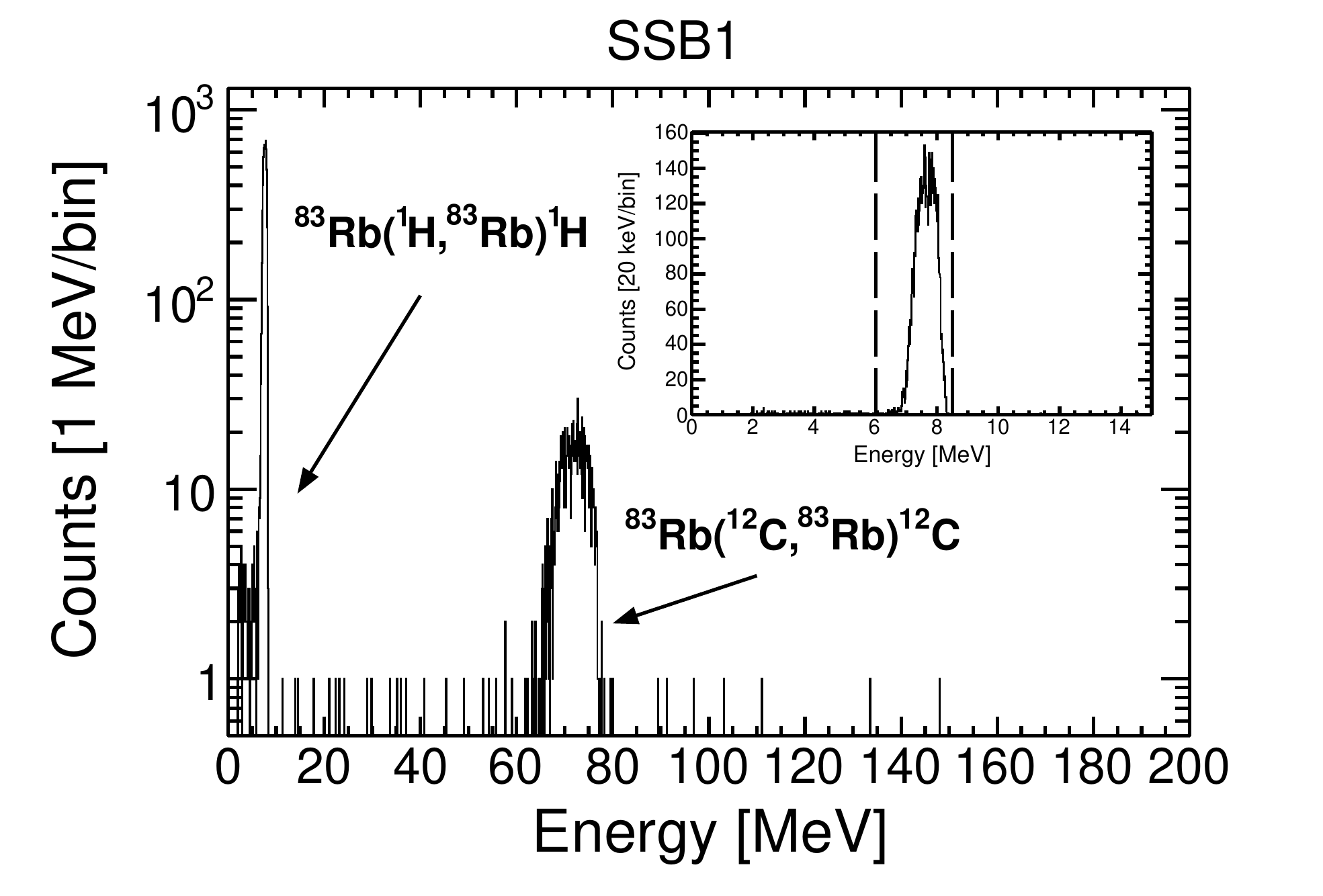}
    \caption{Typical energy spectrum from one of the SSB detectors. The proton and carbon scattering peaks are labelled; the target constituents scattered by the $^{83}$Rb and $^{83}$Sr beam components are indistinguishable on the basis of energy and contribute to both of these peaks. The inset shows the same plot zoomed in on the proton scattering peak with selection cuts indicated by the vertical dashed lines.}
    \label{fig:ssb}
    \end{figure}

Elastically scattered C and H target constituents were detected using two 150 mm$^2$ silicon surface barrier (SSB) detectors located 5~cm downstream of the target and centred at $20^\circ$ angles with respect to the beam axis \cite{EMMA}, allowing for continuous monitoring of the experimental luminosity. The SSBs were fitted with thick Al caps that have central 3~mm diameter apertures to limit the counting rates and to protect the detectors. Protons scattered into these detectors are readily identified by their deposited energy, as indicated in the spectrum shown in Fig.\ \ref{fig:ssb}.
	
The instantaneous rate of proton detections in the SSB detectors is directly proportional to the product of the beam current $I$ and the areal hydrogen number density of the target $n$. This proportionality is expressed via the constant $R$, which is defined by Equation \ref{eqn:rfactor} and calculated using the measured target thickness and data from the first five minutes of each measurement on a fresh target.\begin{equation} \label{eqn:rfactor}
R \equiv \frac{f I}{Q e} \frac{\Delta t}{\Delta N_{p}}n, 
\end{equation}
where $f$ is the fraction of the beam current accounted for by the ion of interest, $Q e$ is the charge of each beam ion, and $\Delta N_{p}$ is the number of scattered protons detected during the 5 minute time interval $\Delta t$. There is a different proportionality constant $R$ for each combination of beam, target, and SSB. The beam current was measured with a relative precision of $\pm10\%$ at 1~h intervals immediately prior to starting each data-taking run; in the case of the $^{84}$Kr beam, this was done using a Faraday cup 1~m upstream of the EMMA target position while for the radioactive $^{83}$Rb beam we used a Faraday cup located 19~m upstream of the target chamber. The transmission from both upstream Faraday cups to the Faraday cup located at the EMMA target position was measured to be 100\%. The integrated luminosity of the yield measurement on each target is given by Equation \ref{eqn:rfactor2}.
\begin{equation} \label{eqn:rfactor2}
\int \mathcal{L}(t) dt = \int \frac{d(N_{b} n)}{dt}dt= R N_{p},
\end{equation}
where $N_b$ is the number of ions of interest incident on the target and $N_{p}$ is the total number of detected protons scattered from the target.

The areal number density of each target was ascertained with a relative precision of $\pm10\%$ prior to the experiment by measuring the energy losses of $\alpha$ particles from a standard triple $\alpha$ source, with stopping powers determined by the computer code SRIM \cite{SRIM}. Table \ref{tab:NbeamNtarget} gives the integrated luminosity for each yield measurement, calculated as the unweighted average of the luminosities found with each SSB detector. 

%In the case of the $2.4 A$~MeV yield measurement, which used two targets of differing thickness, the integrated luminosity was calculated by summing the individual $N_{b} \cdot N_{t}$ over the two targets used. 

%Note that for the radioactive beam, $N_{b}$ is the total beam and must be further multiplied by the $^{83}$Rb beam fraction 
	
	\begin{ruledtabular}
	\begin{table*}[t!]
	    \centering
	    \caption{Target densities and integrated luminosities for the various beams, energies and targets used in this study. The integrated luminosity represents the product of the total number of incident beam ions and the areal target density.}
	    \begin{tabular}{c c c c}
	       Bombarding Energy (MeV) &  Beam   & Target Density ($\mu$g cm$^{-2}$) & Integrated Luminosity ($\mu$b$^{-1}$) \\ \hline
	        $2.7 A$           &  $^{84}$Kr & 727(73) &  $12.1 \pm 0.6_{\mathrm{stat}} \pm 1.7_{\mathrm{sys}}$\\
	        %2.7 & RIB & 900 & $4.59 \pm 0.48_{\mathrm{stat}} \pm 0.66_{\mathrm{sys}}$ \\
	        $2.7 A$ & $^{83}$Rb & 900(90) & $28.3 \pm 3.0_{\mathrm{stat}} \pm 4.3_{\mathrm{sys}}$ \\
	       % 2.4 & $^{83}$Rb & 353 & $1.85 \pm 0.21_{\mathrm{stat}} \pm 0.22_{\mathrm{sys}}$ \\
	        $2.4 A$ & $^{83}$Rb & 353(35) & $11.5 \pm 1.3_{\mathrm{stat}} \pm 1.4_{\mathrm{sys}}$ \\
	        %$2.4 A$ & $^{83}$Rb & 330(33) & $0.73 \pm 0.04_{\mathrm{stat}} \pm 0.10_{\mathrm{sys}}$
	        $2.4 A$ & $^{83}$Rb & 330(33) & $4.5 \pm 0.3_{\mathrm{stat}} \pm 0.6_{\mathrm{sys}}$
	    \end{tabular}
	    \label{tab:NbeamNtarget}
	\end{table*}
	\end{ruledtabular}

\subsection{\label{sec:csd} Recoil Charge State Fractions}
	
\begin{figure}[!t]
    \includegraphics[width=\linewidth]{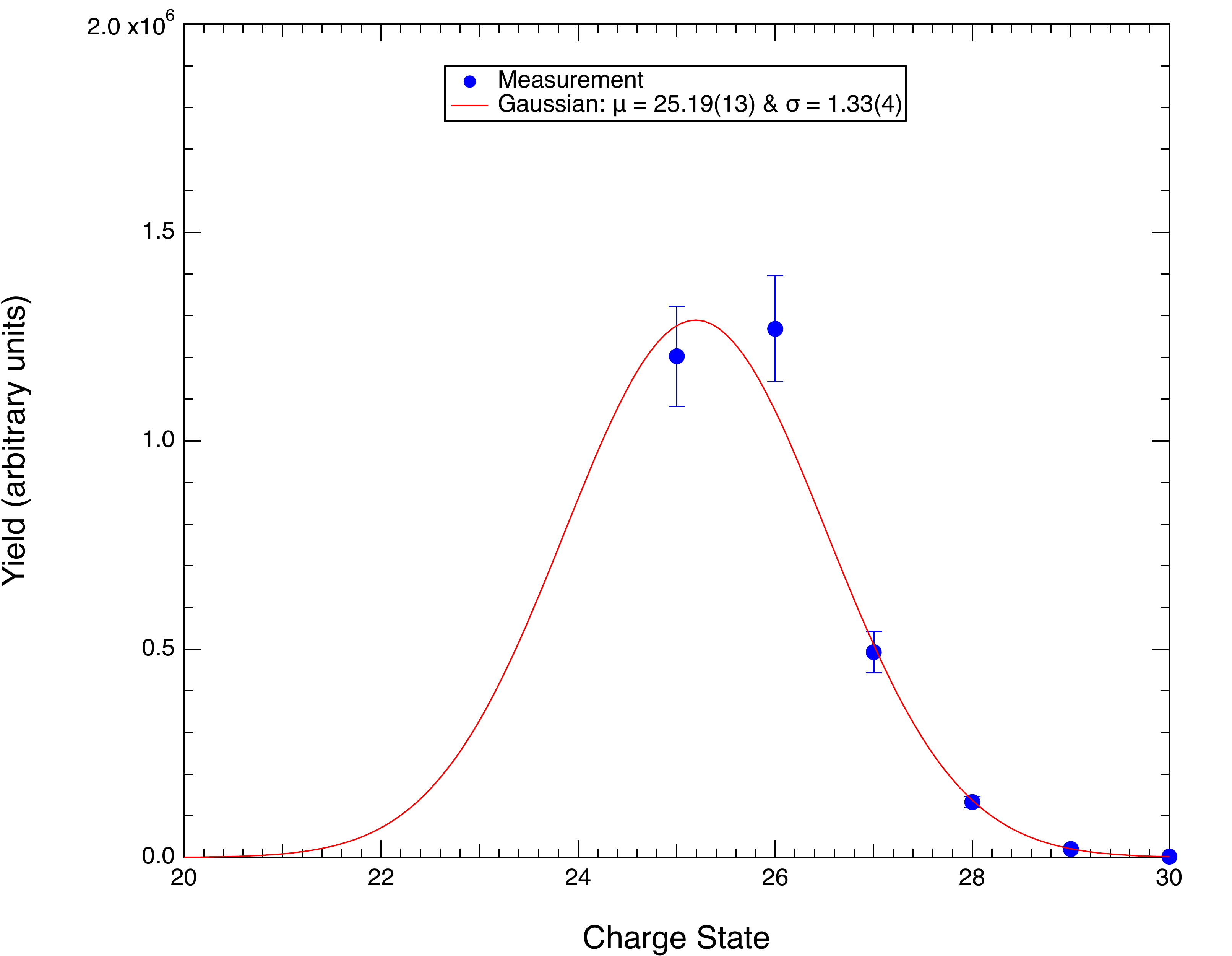}
    \caption{Measured charge state distribution of $^{84}$Kr incident at $2.7 A$~MeV emerging from the 727 $\mu$g~cm$^{-2}$ polyethylene target.  EMMA was set for a kinetic energy of 166 MeV during all of the yield measurements. The error bars represent systematic beam current uncertainties estimated to be $\pm10\%$. Also shown are the results of a Gaussian fit.}
    \label{fig:csd_kr84}
    \end{figure}

In order to optimize the suppression of scattered beam, the focal plane slit system was configured so that only a single charge state of the radiative capture recoils would be transmitted to the focal plane detectors. Therefore, to determine the full reaction yield, the fraction of recoils represented by the selected charge state must be determined for each yield measurement. The charge state distribution of the $2.7A$ MeV $^{84}$Kr beam emerging from the 727 $\mu$g~cm$^{-2}$ target was measured by attenuating its intensity to the order of 1000~s$^{-1}$. This intensity reduction was achieved using wire mesh attenuators and slit systems just downstream of the offline ion source, thereby reducing the intensity without changing the energy or position of the beam on the reaction target. Steady beam current was maintained while 6 charge states of $^{84}$Kr ions were transported successively to the EMMA focal plane and counted over 5 minute intervals. A scintillator located 1~m upstream of the EMMA target position was used to measure the beam intensity before and after each charge state was transmitted. The spectrometer was set for a kinetic energy of 166 MeV throughout the measurements. Yields were normalized according to the number of  incident beam ions. The measured charge state distribution is shown in Fig.\ \ref{fig:csd_kr84}. All statistical errors are smaller than $\pm3\%$. Systematic uncertainties are dominated by contributions due to beam current fluctuations and are estimated to be $\pm10\%$. The data were fit with a Gaussian whose parameters are specified in Fig.\ \ref{fig:csd_kr84}.
	
The parameters derived from the Gaussian fit to the measured $2.7A$ MeV $^{84}$Kr charge state distribution were used to infer the charge state fractions of $^{85}$Rb and $^{84}$Sr recoils after emerging from their respective targets, using the dependence of the mean and standard deviation of the equilibrium charge state on $Z$ and kinetic energy predicted by the empirical parametrization of Ref.\ \cite{shima}. The models of References \cite{nikolaev68,schiwietz01} agree very closely with that of Ref.\ \cite{shima} regarding these dependences. The small differences in kinetic energy and $Z$ of the detected $^{85}$Rb and $^{84}$Sr recoils with respect to those of the transmitted $2.7A$ MeV $^{84}$Kr beam ions imply that the calculated mean and standard deviations of the recoil charge state distributions were larger than the corresponding parameters inferred from the measured charge state distribution by less than 4\% in all cases. Table \ref{tab:CSFs} contains the inferred charge state fractions for the recoils detected in each yield measurement. A relative systematic uncertainty of $\pm10\%$ was adopted for the calculated recoil charge state fractions.

%    \begin{figure}[!t]
  %  \includegraphics[width=\linewidth]{Compare_CSDs_84Kr.pdf}
   % \caption{Comparison between the fit to the $^{84}$Kr CSD data and several semi-empirical model predictions.}
    %\label{fig:csd_kr84_comp}
    %\end{figure}

    \begin{ruledtabular}
    \begin{table*}[t!]
    \centering
     \caption{Charge state fractions for each yield measurement calculated using the parameters inferred from the Gaussian fit to the charge state distribution measured with the $2.7A$ MeV $^{84}$Kr beam and the dependence of the mean and standard deviation of the equilibrium charge state on $Z$ and kinetic energy predicted by the empirical parametrization of Ref.\ \cite{shima}}
    \begin{tabular}{c c c c}
    Recoil & Target Density ($\mu$g cm$^{-2}$) & Selected Charge State ($e$) & Charge State Fraction (\%)    \\ \hline
    $^{85}$Rb  &   727(73)   & 25  &   27.3(27)                    \\
    $^{84}$Sr   &   900(90) &  26  &   27.2(27)                    \\
    $^{84}$Sr   &   353(35) &  26  &   29.3(29)                    \\
    $^{84}$Sr   &   330(33)  &  25  &   21.2(21)                    \\
    \end{tabular}
    \label{tab:CSFs}
    \end{table*}
    \end{ruledtabular}

\subsection{\label{sec:pid}Channel Identification}

%\begin{figure}[!ht]
%\includegraphics[width=\linewidth]{Fig1.pdf}
%\caption{\label{fig1}(a) Time difference between events observed in the TIGRESS $\gamma$-ray array and the focal plane of the EMMA recoil mass spectrometer, following the $^{84}$Kr(p,$\gamma)^{85}$Rb reaction. (b) Energies of $\gamma$ rays observed in coincidence with $A=85$ recoils having TIGRESS-EMMA correlation times falling within the timing peak shown in panel (a).}
%\end{figure}

\begin{figure}[!ht]
\includegraphics[width=\linewidth]{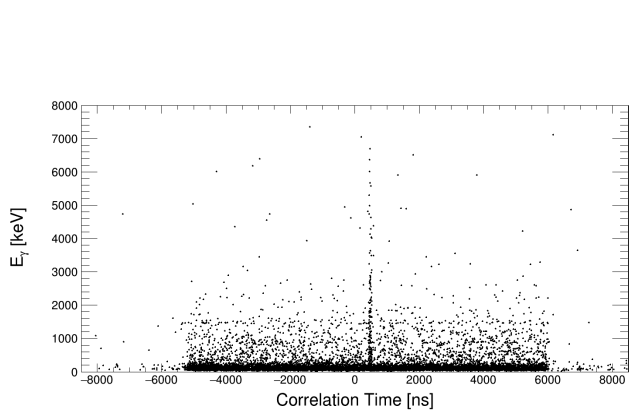}
\caption{\label{84Krgammas}Energies of $\gamma$ rays detected in the TIGRESS array during a measurement of the $^{84}$Kr(p,$\gamma)^{85}$Rb reaction as a function of  TIGRESS-EMMA correlation time. A vertical cluster of counts indicates the observation of correlated primary and secondary $\gamma$ rays, with energies approaching the $p$ emission threshold in $^{85}$Rb of 7~MeV, corresponding to $^{84}$Kr(p,$\gamma)$ events.}
\end{figure}

\begin{figure}[!ht]
\includegraphics[width=\linewidth]{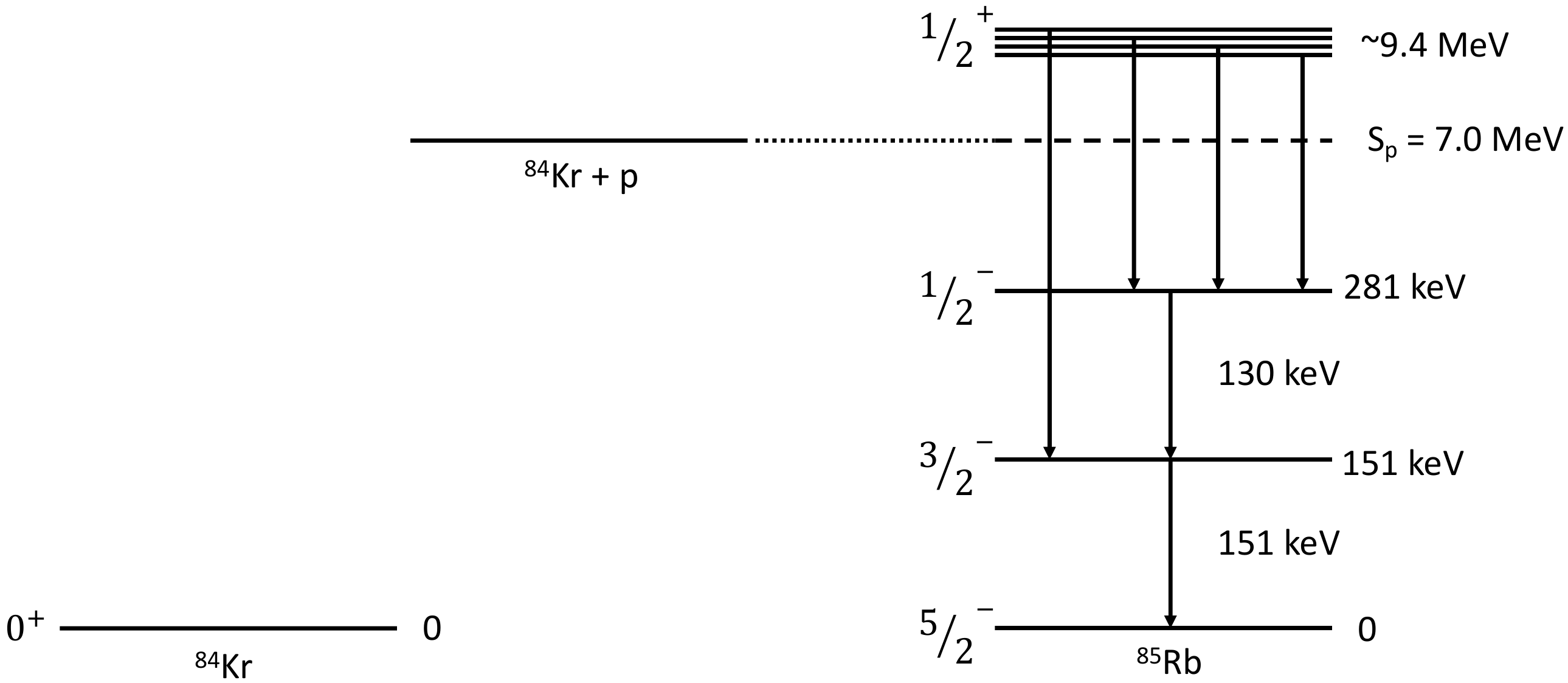}
\caption{\label{85Rb_levels}Simplified energy level diagram of $^{85}$Rb showing only levels relevant to this measurement of the $^{84}$Kr(p,$\gamma)^{85}$Rb reaction. The states initially populated via $s$-wave capture are shown schematically about 2.4~MeV above the proton separation energy $S_p$.}
\end{figure}

A plot of the energies of $\gamma$ rays detected in TIGRESS versus the time difference between $\gamma$-ray events registered in TIGRESS and recoils detected at the focal plane of EMMA is presented in Fig.\ \ref{84Krgammas}. It exhibits a timing peak that provides clear evidence for distinct (p,$\gamma$) events; by placing a software gate on this peak for the measurement of the $^{84}$Kr(p,$\gamma$)$^{85}$Rb reaction, 130- and 151-keV $\gamma$ rays, corresponding to decays from the 1/2$^{-}_1$ and 3/2$^{-}_1$ levels in $^{85}$Rb \cite{Singh2}, were unambiguously identified. In this case, the $1/2^{-}_1$ and $3/2^{-}_1$ excited states were populated following primary $\gamma$ decays from high-lying, proton-unbound levels in $^{85}$Rb. As such, the observed $\gamma$-ray intensities provide direct measures of the inclusive partial reaction cross sections. Note, e.g., that the $1/2^{-}_1$ state decays 99.42(9)\% of the time to the $3/2^{-}_1$ level \cite{Singh2}, so the total radiative capture cross section is not the sum of all the partial cross sections. Rather, the total cross section can be inferred from the measured partial cross section and the calculated branching ratio for $\gamma$-cascade decay through each state. The decay branching ratios of 30$\%$ and 70$\%$ to the 1/2$^{-}_1$ and 3/2$^{-}_1$ excited states in $^{85}$Rb, respectively, predicted by a simplified $\gamma$-cascade model, are expected to be accurate to within $\pm$10$\%$. A simplified $^{85}$Rb level scheme is shown in Fig.\ \ref{85Rb_levels}.

\subsection{\label{sec:eff} Detection Efficiencies}

The $\gamma$-ray detection efficiencies were established using standard $^{152}$Eu and $^{56}$Co sources. Estimates of the relative uncertainties associated with the integrated luminosity, the recoil transmission efficiency, $\gamma$-ray detection efficiency, and charge state fractions amount to $\pm19\%$, $^{+0.1}_{-33}$\%, $\pm5\%$, and $\pm10$\%, respectively. We note that the recoil transmission efficiency is believed to be high based on the small recoil cone angle and small kinetic energy spread of $\pm1$\%. However, we have estimated its downward uncertainty conservatively to account for the possibility of unforeseen losses during the measurement of the (p,$\gamma$) reaction cross sections due to uncertainties in the stopping power of the recoils in the thick targets. The transmission efficiency through the spectrometer is a function of the scattering angle and the relative kinetic energy per charge deviation of the recoil with respect to that of the reference trajectory for which the spectrometer has been set. On account of the small sizes of the maximum scattering angle due to the kinematics of the reaction and the additional deflection resulting from multiple scattering in the target, the uncertainties associated with calculating the latter do not have a substantial effect on the estimated recoil transmission efficiency of 99.5\%. However, even at these small angles with respect to the optic axis, the transmission efficiency of the spectrometer with these restrictive slit settings nearly vanishes for recoil relative kinetic energy/charge deviations beyond $\pm12$\%. The target thicknesses were measured to a relative precision of $\pm10$\% and we estimate the stopping power uncertainties to be $\pm5$\% for $^{83}$Rb in polyethylene at these energies, resulting in a $\pm4.2$\% uncertainty in the calculated recoil kinetic energy/charge for the $2.7A$~MeV measurement. This uncertainty in the kinetic energy/charge leads to a large downward uncertainty on the recoil transmission efficiency estimated to be $-33$\%, while the upward uncertainty of 0.1\% is much smaller since the transmission efficiency cannot be greater than 1. The data acquisition live-time fraction exceeded 90\% for data taking with both beams and has a negligible statistical uncertainty.

%\subsection{Theoretical Methods}

\subsection{\label{sec:effene}Effective Energy}

For the measurements of the $^{84}$Kr(p,$\gamma$)$^{85}$Rb and $^{83}$Rb(p,$\gamma$)$^{84}$Sr reactions, the effective relative kinetic energy, $E_{\mathrm{cm}}$, was determined from the incident beam energy and energy loss through the (CH$_2$)$_n$ target, assuming a reaction cross section energy dependence similar to the one obtained from statistical model calculations
\cite{NONSMOKER1,NONSMOKER2}. Specifically, effective energies were calculated by solving Equation \ref{eq1} for $E_{\mathrm{cm}}$.
\begin{equation}\label{eq1}
\langle \sigma(E) \rangle=\frac{\int_{E_f}^{E_i} \sigma(E) dE}{\int_{E_f}^{E_i} dE}=\sigma(E_{\mathrm{cm}})
\end{equation}
The energy loss of the beam $E_i-E_f$ was calculated using the program LISE++ \cite{LISE}. It employs SRIM stopping powers, which are assumed to be known to $\pm3.9$\% for $^{84}$Kr and $\pm$5\% for $^{83}$Rb. The uncertainty in the effective energies includes a contribution due to the stopping powers, a contribution from the target thickness, and one from the uncertainty in the energy dependence of the cross section. The last of these is estimated via the difference between the effective energy deduced assuming the statistical model and assuming an energy-independent astrophysical $S$ factor.

\section{\label{sec:results}Results}

We observe 22(5) counts due to the 151-keV $\gamma$-ray transition in $^{85}$Rb, resulting from the  $^{84}$Kr(p,$\gamma$) reaction, while 11(4) counts are observed from the 130-keV transition that dominates the decay of the 281-keV state. Combining these yields with the predicted branching ratios in a weighted average, we infer a total reaction cross section at $E_{\mathrm{cm}}$ = 2.443(22) MeV of $133^{+91}_{-44}$ $\mu$b. A summary of the 
parameters used for the determination of the reaction cross sections is given in Table \ref{table}. Due to small differences in the energy loss and charge state fraction calculations which affect the recoil transmission efficiency, the effective energies, detection efficiencies, and cross sections differ slightly from those given in Ref.\ \cite{lotay21}, though they are consistent. The values given here supersede those in our prior work. The inferred total $^{84}$Kr(p,$\gamma$) cross section is smaller than but compatible with the measurements reported in Ref.\ \cite{palmisano21} at nearby energies.

\begin{ruledtabular} 
\begin{table*}[t]
\caption{Parameters used for the determination of partial radiative capture cross sections. The detection efficiency is the product of the recoil transmission efficiency, the recoil charge state fraction, the focal plane detection efficiency, the live-time fraction, and the $\gamma$-ray detection efficiency. Errors are specified at the 68\% CL while upper limits are specified at the 90$\%$ CL. Predicted partial cross sections are based on a statistical model of the reaction and subsequent $\gamma$-ray cascade \cite{SMARAGD}.}

\begin{tabular}{ccccccccc}
 
 Reaction &
 $E_{\gamma}$ &
 Transition &
Integrated& 
Events & 
Detection& 
$E_{\mathrm{cm}}$  &
Measured &
Predicted\\
 & & &Luminosity&&Efficiency&&$\sigma_{\mathrm{partial}}$ &$\sigma_{\mathrm{partial}}$\\
  & (keV) &  & ($\mu$$b$$^{-1}$)  & & ($\%$) & (MeV) & ($\mu$$b$) & ($\mu$$b$) \\ \hline

$^{83}$Rb(p,$\gamma$)$^{84}$Sr & 793 & 2$^{+}$ $\rightarrow$ 0$^{+}$  & 28(5) & 16(6) & 1.1$^{+0.1}_{-0.4}$ & 2.386(23) & 52$^{+40}_{-22}$ & 181(26) \\ 
 & 793 & 2$^{+}$ $\rightarrow$ 0$^{+}$ & 16(2) & $< 16$ & 1.1$^{+0.1}_{-0.4}$ & 2.260(7) & $< 103$ & 110(16) \\ 
 
$^{84}$Kr(p,$\gamma$)$^{85}$Rb  & 151 & 3/2$^{-}$ $\rightarrow$ 5/2$^{-}$ & 12(2) & 22(5) & 2.2$^{+0.3}_{-0.8}$ & 2.443(22) & 83$^{+56}_{-26}$ & 257(40)\\ 
& 130 & 1/2$^{-}$ $\rightarrow$ 3/2$^{-}$ & 12(2) & 11(4) & 2.1$^{+0.3}_{-0.8}$ & 2.443(22) & 44$^{+31}_{-17}$ & 106(40)

\end{tabular}
\label{table}

\end{table*}
\end{ruledtabular}

%\begin{figure}[!ht]
%\includegraphics[width=\linewidth]{Fig2.pdf}
%\caption{\label{fig2}Observed $\gamma$-ray energies in the TIGRESS array, during the measurement of the $^{83}$Rb(p,$\gamma$) reaction, as a function of  TIGRESS-EMMA correlation time. A vertical cluster of counts indicates the observation of correlated primary and secondary $\gamma$ rays up to high energies, corresponding to $^{83}$Rb(p,$\gamma$) events.}
%\end{figure}

In the measurement of the astrophysically important $^{83}$Rb + $p$ reaction, clearly correlated $\gamma$ rays, extending to high energies, were observed at an effective energy of $E_{\mathrm{cm}}$ = 2.386(23) MeV. These events indicate the population of proton-unbound levels in $^{84}$Sr and represent conclusive evidence for the observation of radiative proton captures by $^{83}$Rb. However, there is significant background throughout the low-energy part of the spectrum, due to the $\beta$-delayed $\gamma$ decay of the known isobaric beam contaminant $^{83}$Sr. Nevertheless, it is possible to accurately account for this background using well-known $^{83}$Sr decay data \cite{mccutchan} and by only investigating $\gamma$-decay transitions detected in the 8 detectors centred at 90$^{\circ}$ with respect to the beam axis. In this regard, when applying a Doppler correction appropriate for $^{84}$Sr recoils, $\beta$-delayed transitions from the decays of stopped $^{83}$Sr beam contaminants are shifted into several distinct peaks according to the angles of the detectors, while prompt (p,$\gamma$) transitions are observed as a peak at a single energy. 

\begin{figure*}[!ht]
\includegraphics[width=\linewidth]{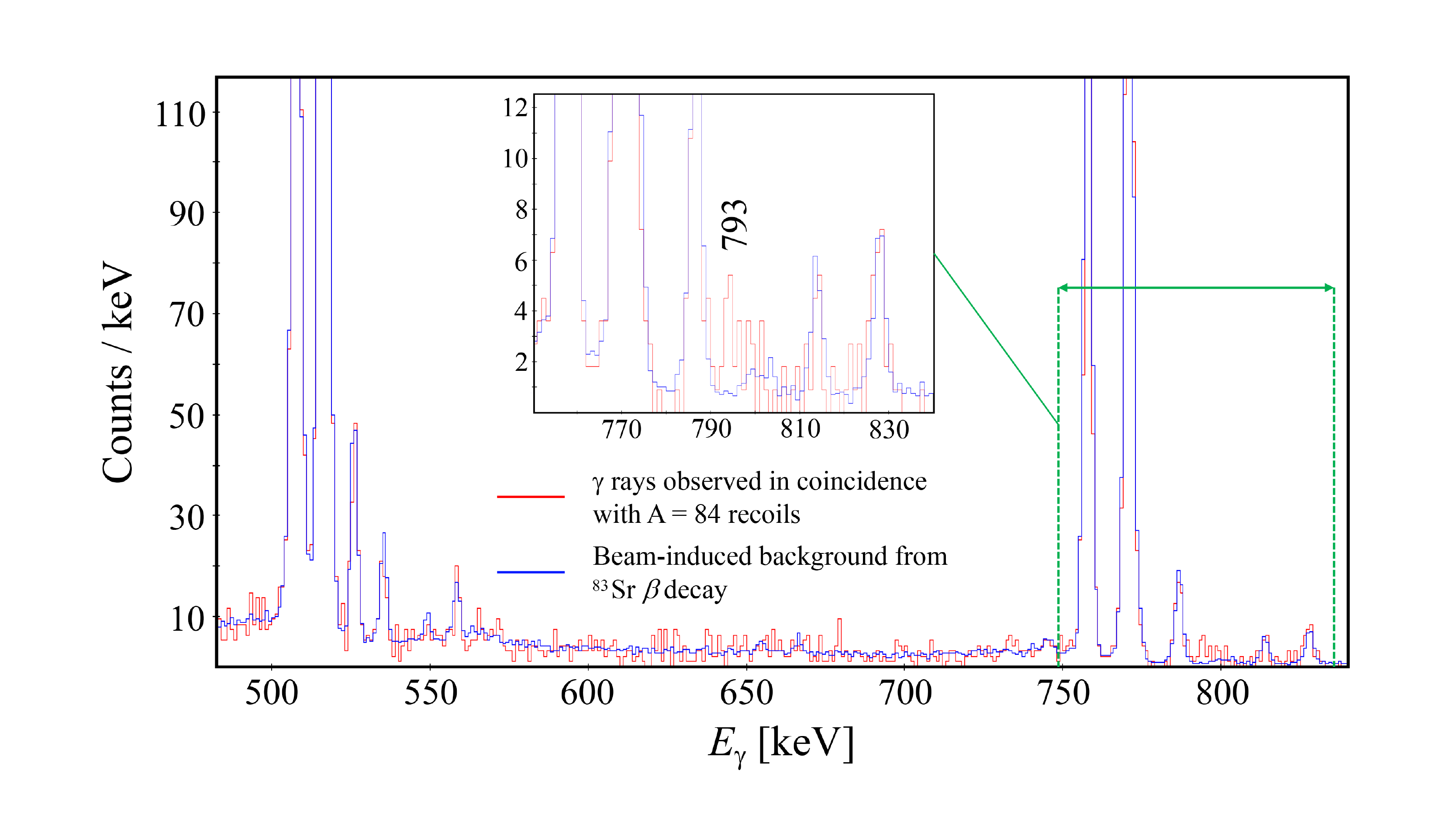}
\caption{\label{83Rbgammas} Gamma rays observed in the 8 TIGRESS detectors centred at 90$^{\circ}$ with respect to the beam axis in coincidence with $A=84$ recoils, following the $^{83}$Rb(p,$\gamma)^{84}$Sr reaction. The inset shows a zoomed-in view of the same spectrum centred about the energy of the 793~keV transition.}
\end{figure*}

\begin{figure}[!ht]
\includegraphics[width=\linewidth]{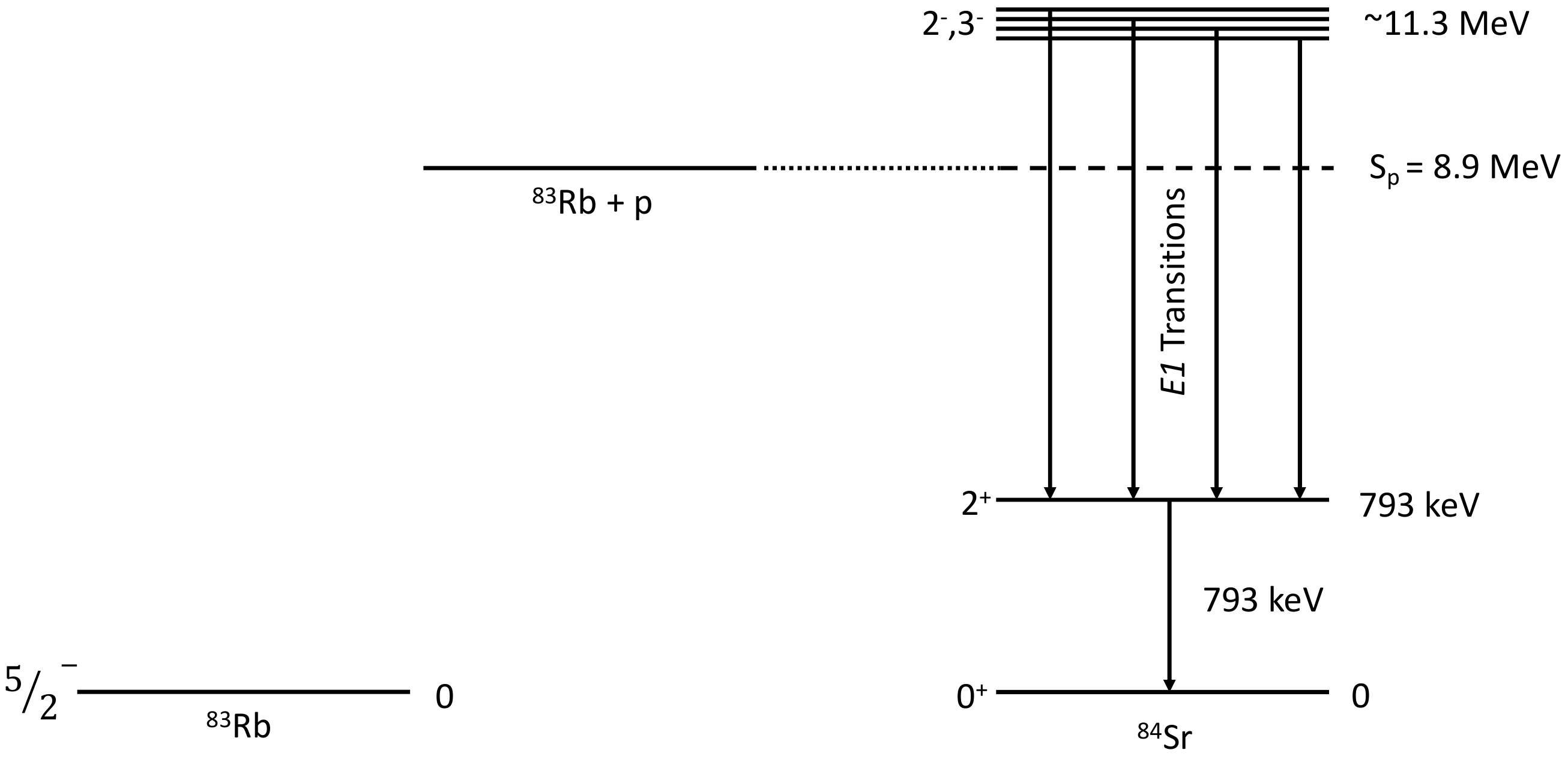}
\caption{\label{84Sr_levels}Simplified energy level diagram of $^{84}$Sr showing only levels relevant to this measurement of the $^{83}$Rb(p,$\gamma)^{84}$Sr reaction. The states initially populated via $s$-wave capture are shown schematically about 2.4~MeV above the proton separation energy $S_p$.}
\end{figure}

Fig.\ \ref{83Rbgammas} illustrates the $\gamma$ decays observed in the 8 TIGRESS detectors centred at 90$^{\circ}$ with respect to the beam axis in coincidence with $A=84$ recoils transmitted to the focal plane of EMMA, during the measurement of the $^{83}$Rb(p,$\gamma$) reaction at $E_{\mathrm{cm}}$ = 2.386 MeV. Here, 16(6) counts, in excess of those expected as a result of beam-induced background, are observed at 793 keV, indicating strong population of the 2$^{+}_1$ excited level in $^{84}$Sr \cite{Singh}. Based on statistical model calculations, it is expected that 70(10)$\%$ of the radiative captures proceed through this state and, in the present work, no other decay branches were observed. As such, we measured the partial cross section to the 2$^{+}_1$ excited state and infer a total radiative capture cross section of 73$^{+57}_{-33}$ $\mu$$b$. A schematic $^{84}$Sr level scheme is shown in Fig.\ \ref{84Sr_levels}.

A second measurement of the $^{83}$Rb(p,$\gamma)^{84}$Sr reaction was performed at $E_{\mathrm{cm}}$ = 2.260(7) MeV. Unfortunately, only a small excess of 6 events above the mean background of 23 was observed in the region of interest at $793\pm3$ keV in the resultant $\gamma$-ray spectrum, corresponding to population of the 2$^{+}_{1}$ excited state in $^{84}$Sr. Therefore, an upper limit was placed on the $^{83}$Rb(p,$\gamma)^{84}$Sr reaction cross section at $E_{\mathrm{cm}} = 2.260$~MeV. This upper limit on the signal in the presence of expected background events was derived using the method of Feldman and Cousins \cite{Feldman}, leading to a limit of $< 16$ $\gamma$-gated, $A=84$ recoils at the 90$\%$ confidence level (CL).

\section{Discussion}
\label{sec:discussion}

\subsection{Comparison to Reaction Theory}
\label{sec:reactheory}

\begin{figure}[t]
\includegraphics[width=\linewidth]{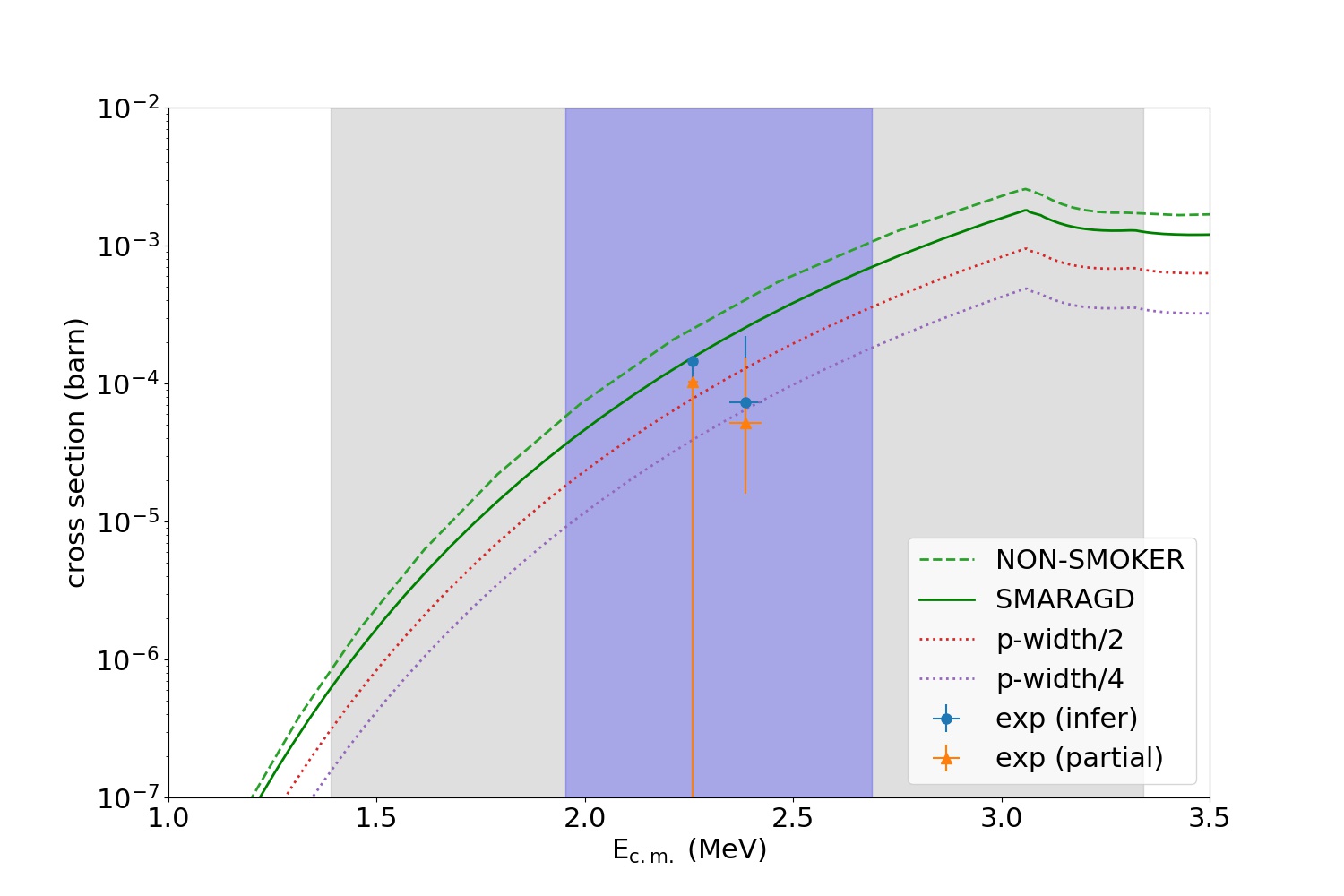}
\caption{Cross section of the $^{83}$Rb(p,$\gamma)^{84}$Sr reaction from experiment (shown are the partial cross section for populating the $2^+_1$ state and the inferred total cross section with 90\% CL error bars) compared to statistical model predictions of the total cross sections with the NON-SMOKER \cite{NONSMOKER2} and SMARAGD \cite{SMARAGD} codes. The wider, lightly shaded region indicates the approximate location of the relevant energy window 
\protect\cite{Rauscher3} for the $^{83}$Rb(p,$\gamma)^{84}$Sr reaction in ccSNe ($2~\mathrm{GK}<T<3.5~\mathrm{GK}$). The narrower, darkly shaded region
indicates the range of relative kinetic energies covered in the measurement of the $E_\mathrm{cm}=2.386$~MeV data point. The measured and inferred points at $E_\mathrm{cm}=2.26$~MeV are 90\% CL upper limits. Also shown are SMARAGD calculations with the default proton width divided by two and four, respectively.}
\label{fig:csexpth83}
\end{figure}

\begin{figure}[t]
\includegraphics[width=\linewidth]{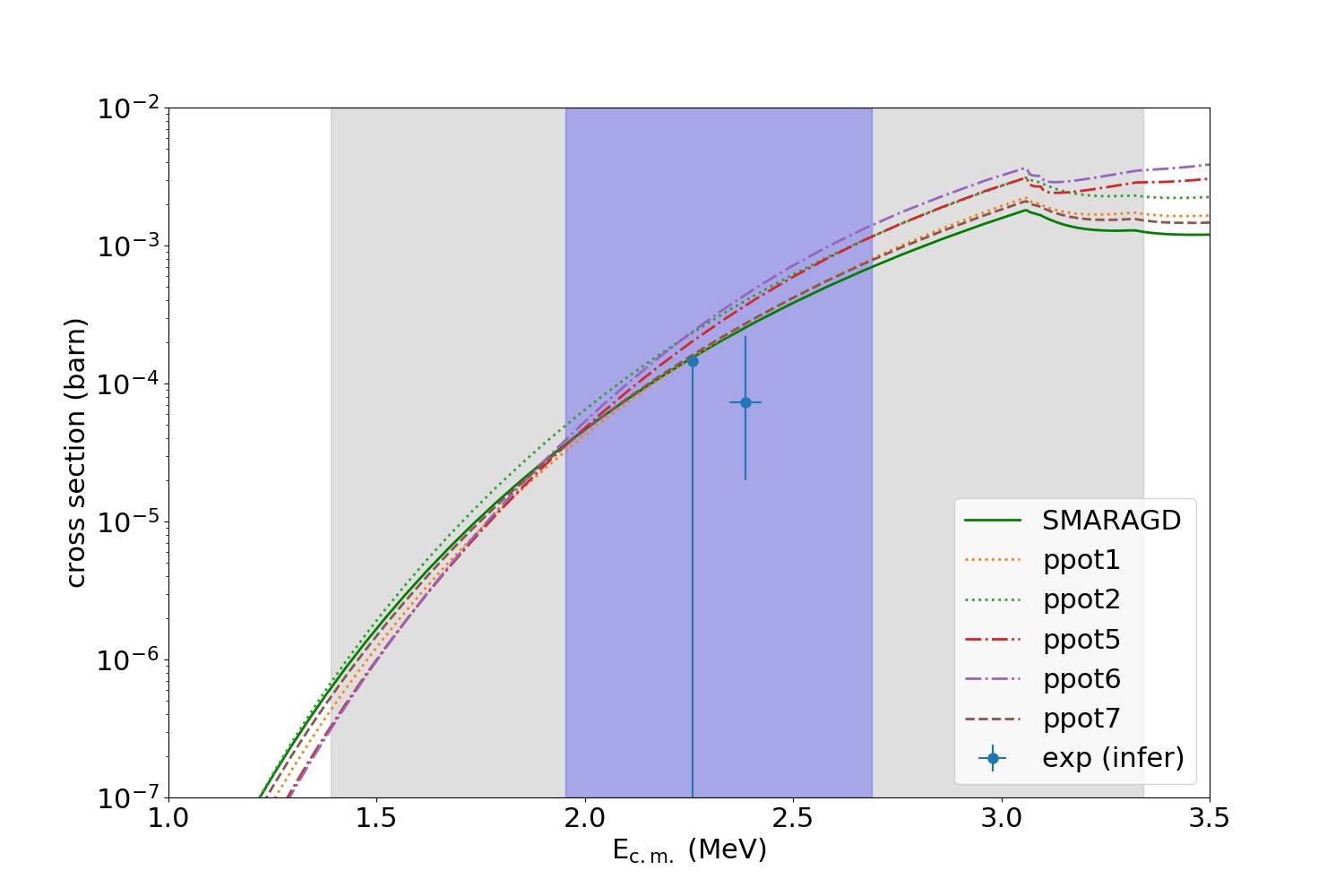}
\caption{Same as Fig.\ \ref{fig:csexpth83} but comparing the experimental data to SMARAGD calculations using various $p+^{83}$Rb optical potentials (SMARAGD default \cite{jlm,lej}, ppot1 \cite{esw}, ppot2 \cite{perey}, ppot5 \cite{baugelane}, ppot6 \cite{bauge}, ppot7 \cite{modjlm}).}
\label{fig:csexpth83pot}
\end{figure}

\begin{figure}[t]
\includegraphics[width=\linewidth]{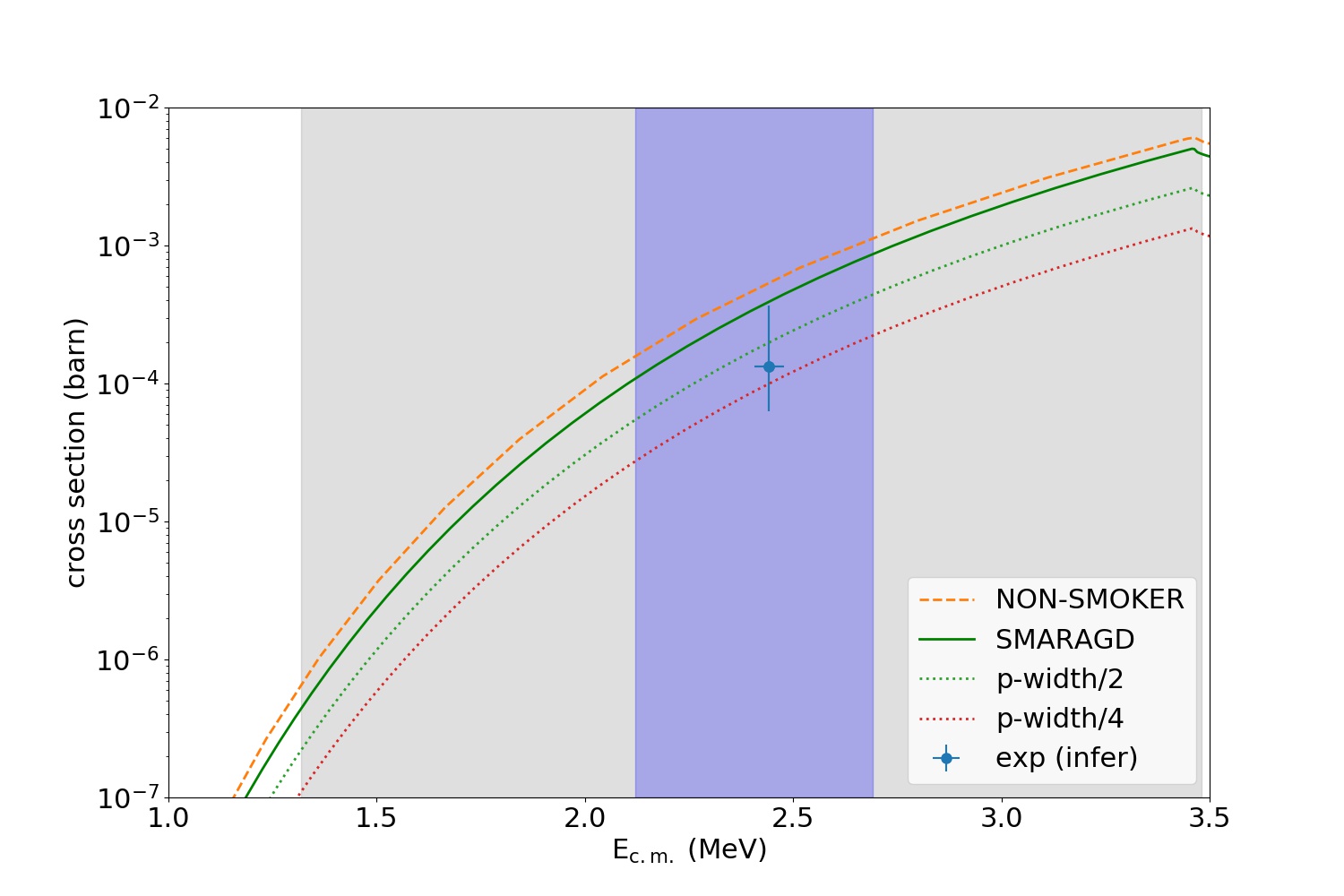}
\caption{Same as Fig.\ \ref{fig:csexpth83} but for the $^{84}$Kr(p,$\gamma$)$^{85}$Rb reaction.}
\label{fig:csexpth84}
\end{figure}

Figures \ref{fig:csexpth83} and \ref{fig:csexpth84} compare 
the ground-state cross sections
predicted by the Hauser-Feshbach (HF) statistical model code NON-SMOKER \cite{NONSMOKER1,NONSMOKER2} to the total cross sections 
inferred from the 
experimentally measured partial $^{83}$Rb(p,$\gamma$) and $^{84}$Kr(p,$\gamma$) reaction cross sections, respectively. In
most astrophysical investigations, the NON-SMOKER results for a wide range of nuclides 
provide the default set of reaction rates in the absence of experimental data. It is difficult to 
use the upper limit of the $^{83}$Rb(p,$\gamma$) cross section at $E_{\mathrm{cm}}$ = 2.260 MeV for an improved prediction but the experimental 
value at $E_{\mathrm{cm}}$ = 2.386 MeV as well as the one for $^{84}$Kr(p,$\gamma$) indicate cross sections smaller
than the NON-SMOKER predictions by roughly a factor of six.

To further understand the source of the difference between the prediction and the data it is necessary to investigate the sensitivity of the cross section to a variation of nuclear properties included in the calculation of the cross section. Such sensitivities were explored in Ref.\ \cite{sensi}. For the present reactions, it was found that, among the $\alpha$, neutron, proton, and radiative widths entering the Hauser-Feshbach calculation, the cross section below the Coulomb barrier and at the measured energies is predominantly determined by the average proton width (as predicted by theory).

We have performed exploratory calculations to assess the required changes to reproduce the experimental cross sections, using the SMARAGD code \cite{SMARAGD}. This code is a further development of the NON-SMOKER code, including more recent nuclear data but also improved theoretical treatments of nuclear properties and improved numerical procedures. 

The proton widths are mainly determined by the $p+^{83}$Rb and $p+^{84}$Kr optical potentials. To a lesser extent, they depend on the number and quantum properties of the states reached in proton emission from the compound nucleus, i.e., energetically accessible excited states in the beam nucleus. For the cases considered here, these states are fairly well known and therefore the optical potential remains the most significant source of uncertainty. Nevertheless, the SMARAGD cross section is lower by about 30\% than the NON-SMOKER cross section even when using the same default optical potential of Refs.\ \cite{jlm,lej}. This is due to a different numerical approach to solving the Schr\"odinger equation to compute wave functions and charged-particle transmission coefficients. The improved method used in the SMARAGD code is superior at sub-Coulomb energies and leads to the reduction relative to the NON-SMOKER prediction seen in Figs.\ \ref{fig:csexpth83} and \ref{fig:csexpth84}. This reduction causes the standard SMARAGD value to be close to the experimental 90\% CL region. As is also shown in Figs.\ \ref{fig:csexpth83} and \ref{fig:csexpth84}, a proton width approximately 0.3 times as large as the width predicted by SMARAGD would reproduce the experimental cross section inferred to be most likely.

\begin{figure}[t]
\includegraphics[width=\linewidth]{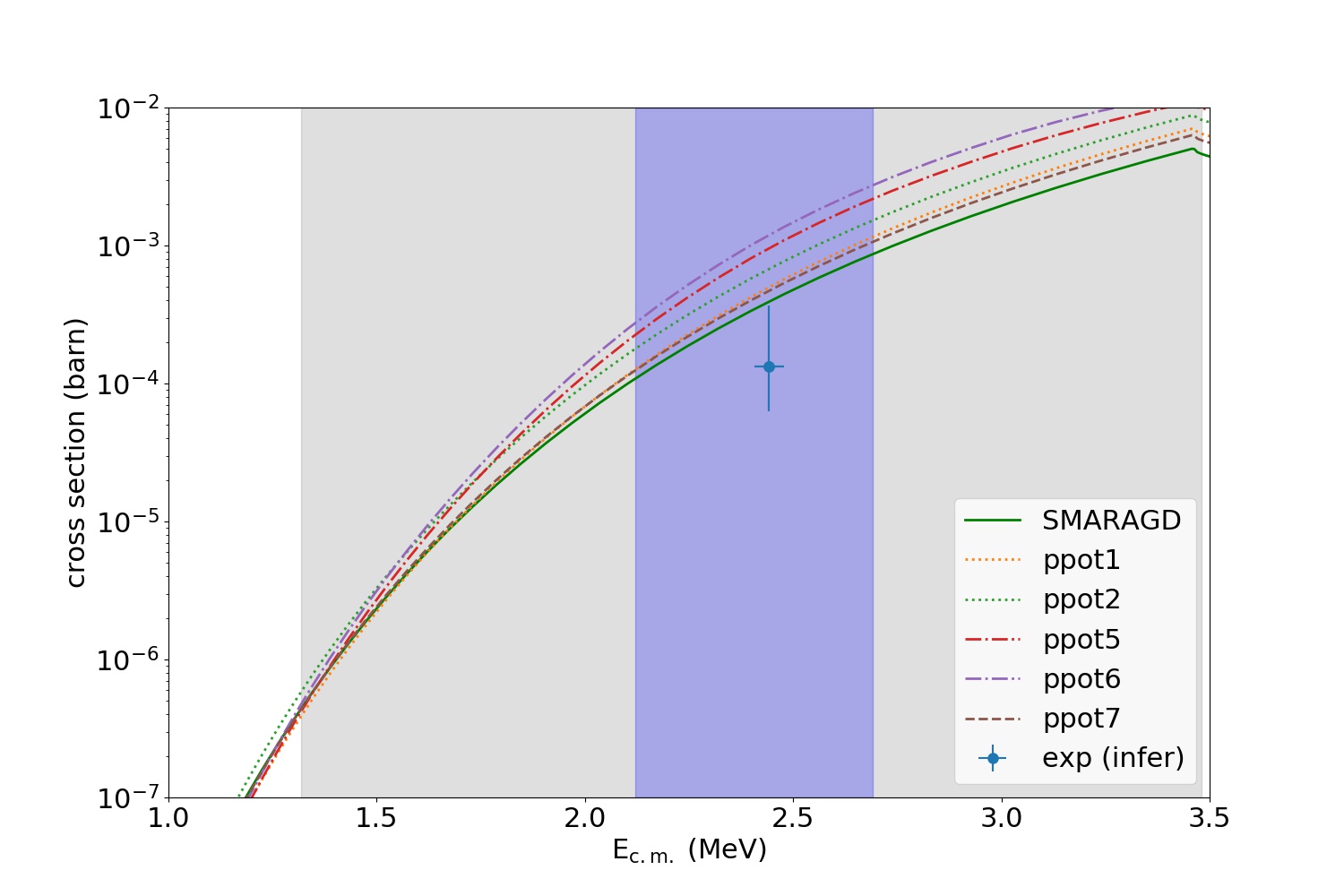}
\caption{Same as Fig.\ \ref{fig:csexpth83pot} but for the $^{84}$Kr(p,$\gamma$)$^{85}$Rb reaction.}
\label{fig:csexpth84pot}
\end{figure}

In order to estimate the uncertainty connected to the use of the optical potential, we have performed calculations with additional optical potentials taken from literature: a simple equivalent square-well potential (ppot1, \cite{esw}), a Saxon-Woods parameterization with energy- and mass-dependent parameters (ppot2, \cite{perey}), a re-parameterization of the potential of Ref.\ \cite{jlm} based on more recent data (ppot5, \cite{bauge}) and a Lane-consistent version of this (ppot6, \cite{baugelane}). Additionally, a recent modification of the default microscopic potential of References \cite{jlm,lej} that has provided an improved description of low-energy data in the $A\approx 80$ mass range with an increased imaginary part (ppot7, \cite{modjlm}) was used. As can be seen in Figs.\ \ref{fig:csexpth83pot} and \ref{fig:csexpth84pot}, respectively, all these potentials lead to even larger cross sections at the measured $E_{\mathrm{cm}}$ than the default SMARAGD calculation, which is the most consistent with the experimental values.

It is to be noted that due to the low energy, the penetration through the Coulomb barrier dominates the transmission coefficients and the actual shape of the imaginary potential is of lesser importance. An independent investigation using a simple barrier penetration model with a real potential  corroborated the results obtained with the optical potential approach and likewise was unable to obtain cross sections small enough to match the central experimental values \cite{mohrprivcomm}. Moreover, previous studies of low-energy $(p,\gamma)$ and $(p,n)$ reactions on stable targets with masses $A>70$ have not seen such large discrepancies yet (see, e.g., Refs.\ \cite{Rauscher2,rauintjmodphys} and references therein).

%\subsection{\label{sec:astro_imp}Astrophysical Implications}
\subsection{Thermonuclear Reaction Rate and Astrophysical Implications}
\label{sec:astro}

The determination of a thermonuclear reaction rate for use in astrophysical simulations requires the knowledge of the cross 
sections across the relevant energy range for which an integration over the cross section folded with the energy 
distribution of the protons in a stellar plasma is performed. Plasma temperatures for modifying abundances through a $\gamma$ process in 
stellar explosions range from 2 GK to 3.5 GK, which translates to about $E_{cm}=1.4-3.3$ MeV for the 
$^{83}$Rb(p,$\gamma$) reaction \cite{Rauscher3}. Among the reactions experimentally investigated here, only the $^{83}$Rb(p,$\gamma$) reaction is of astrophysical significance, as discussed below. 
Moreover, proton-induced reactions on the ground state of $^{83}$Rb contribute only about 20-30\% of the stellar reaction rate 
\cite{sensi}. This is due to the fact that, in an astrophysical plasma at $2-3.5$ GK a large fraction of the $^{83}$Rb nuclei are present in thermally excited 
states. So far, the contributions of excited states can only be treated by theory \cite{Rauscherb}. As a consequence, a 
measurement with a beam in the ground state at one energy is not sufficient in itself to fully constrain the astrophysical 
reaction rate, even when the energy is within the astrophysically relevant energy range. As discussed earlier (in Sec.\ \ref{sec:reactheory}), however, the data determine the
ground-state cross sections and, when compared to statistical model predictions, help to constrain certain reaction properties also important in reactions on excited target states.

Therefore, in order to evaluate the impact of the present measurement on astrophysical simulations of the $\gamma$ 
process, we constructed the stellar reaction rate by multiplying the standard rate used in the simulation by the ratio of the experimental and predicted cross sections at $E_{\mathrm{cm}}$ = 2.386 MeV, 
which is roughly one sixth for the NON-SMOKER reaction rate previously used \cite{NONSMOKER1}. Although we have highlighted some of the difficulties associated with theoretically calculating such a small cross section, which is not required for consistency with the experiment, we chose this value to investigate the largest possible impact on the astrophysical result. Reducing the \textit{stellar} reaction rate by the same factor as the ground state cross section further implies
that the excited state contributions require the same renormalization as the ground state cross section, which is also the most extreme case \cite{2012ApJ...755L..10R}. Further experimental studies comparing the actual energy dependence of the cross section to the predicted energy dependence would be required to judge the validity of this assumption.

The impact of a single reaction in an astrophysical context is often discussed by showing how strongly the abundance of a given 
nuclide changes when varying the reaction rate by a given amount. Although this may provide clues on the general sensitivity of the 
abundance to the rate, it is not wholly sufficient to assess the actual astrophysical impact in an environment where a large number of 
reactions, each with their individual uncertainties, conspire to yield the abundance of a nuclide. In the assessment of the 
importance of a reaction in an ensemble of many reactions, the sensitivity of an abundance to a rate cannot be decoupled a priori 
from the size of uncertainty because a rate with a large uncertainty and a small abundance sensitivity may contribute more to the total 
abundance uncertainty than a rate with a small uncertainty and a large abundance sensitivity \cite{2018AIPC.1947b0015R}. This is 
especially true for the production of $p$ nuclides in a $\gamma$ process.

The recent studies of Refs.\
\cite{Rauscher,nobsnIa} addressed the question of which reactions dominate the uncertainties of $p$-nuclide abundances in core-collapse supernovae and in thermonuclear supernovae, respectively. They 
identified key reactions giving rise to the largest uncertainties in abundances of $p$ nuclides by applying a Monte Carlo (MC) variation 
to a large set of reaction rates within their theoretical or experimental uncertainties. Although $^{83}$Rb(p,$\gamma$) was not 
identified as a key reaction, with its uncertainty solely dominating the abundance uncertainty of a $p$ nuclide, it was found to 
significantly contribute to the uncertainty in the predicted abundance
of $^{84}$Sr in core-collapse supernovae
(see Table 8 in Ref.\ \cite{Rauscher}). The $^{84}$Sr abundance was found to be anti-correlated with the 
$^{83}$Rb(p,$\gamma$) reaction rate.

Here, we follow the same approach as in Refs.\ \cite{Rauscher,nobsnIa}, using the same standard rate library and the same uncertainties except 
for the rate of the $^{83}$Rb(p,$\gamma$)$^{84}$Sr reaction and its inverse, $^{84}$Sr($\gamma$,p)$^{83}$Rb. For the 
$^{83}$Rb(p,$\gamma$)$^{84}$Sr reaction rate we used the renormalized standard rate as described above. Since the thermally averaged rates of a reaction and its inverse are connected by the detailed balance theorem \cite{Rauscherb}, the rate of the  
$^{84}$Sr($\gamma$,p)$^{83}$Rb reaction is renormalized by the same factor. Reaction network calculations were performed for the mass zones of a 15 and a 25 
$M_\odot$ star with solar metallicity, as obtained from the stellar model code KEPLER (see Ref.\ \cite{Rauscher} for details), and for a double-detonation model of a Chandrasekhar-mass White Dwarf (model DDT-a of Ref.\ \cite{nobsnIa}).

\begin{figure}
\includegraphics[width=\columnwidth]{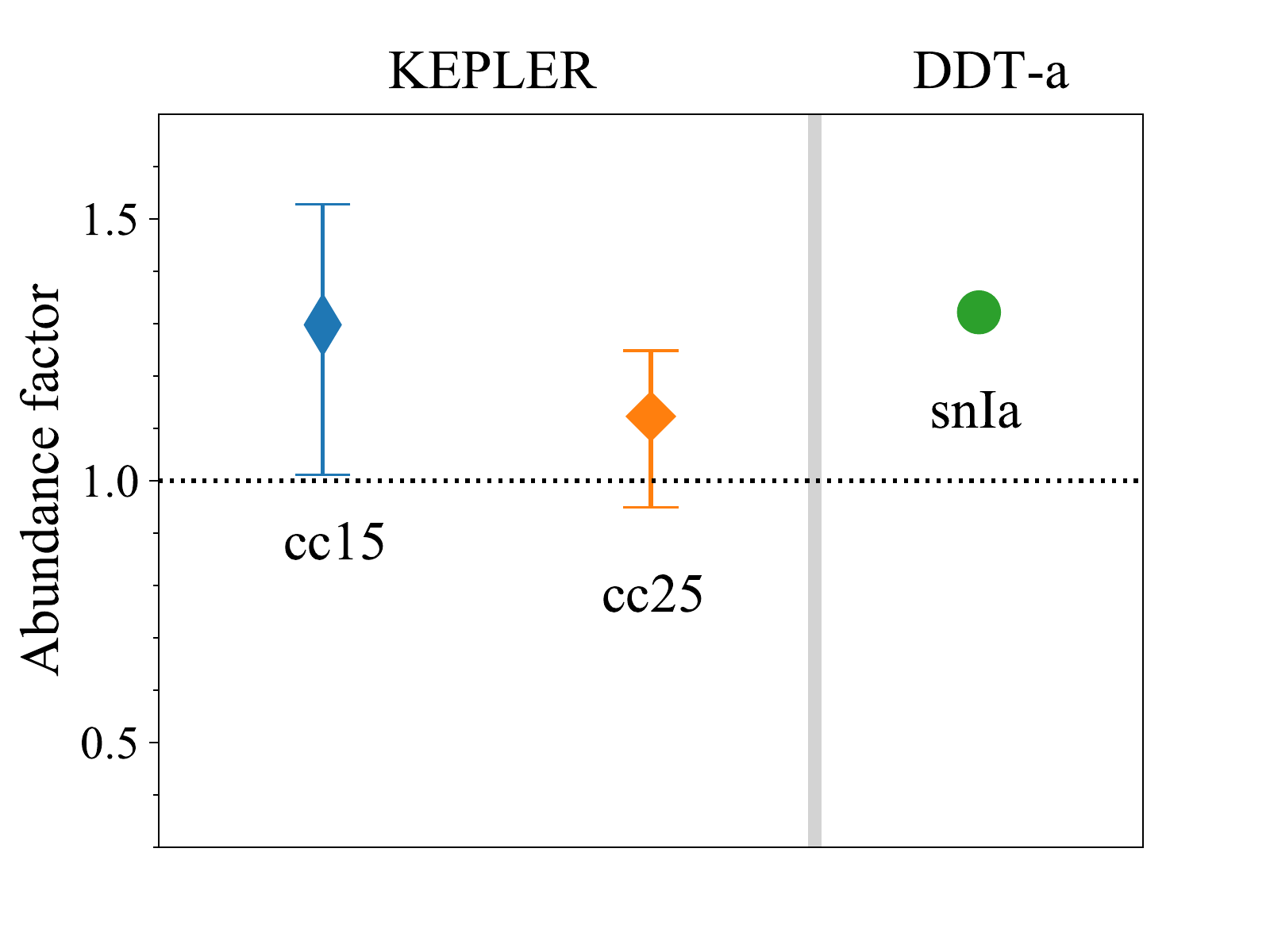}
\caption{Relative change in the abundance of $^{84}$Sr in a 15 (cc15) and 25 (cc25) $M_\odot$ star with solar metallicity exploding as a core-collapse supernova, and in a thermonuclear supernova (snIa),
when using the new rates for $^{83}$Rb(p,$\gamma$)$^{84}$Sr and its inverse reaction. The error bars illustrate the remaining uncertainties due to the combined effect of all reaction rates (see text for details).\label{fig:ccabuns}}
\end{figure}

Fig.\ \ref{fig:ccabuns} shows the change in the $^{84}$Sr abundance for 
the three supernova models obtained when replacing the previously used rates by the rates derived from the present experiment.
%The production of $^{84}$Sr is increased by 27\%, 11\%, and 33\% for the cc15, cc25, and sn Ia models, respectively, due to the reduction of the $^{84}$Sr($\gamma$,p)$^{83}$Rb rate.
%% update by NN
The production of $^{84}$Sr is increased by 30\%, 12\%, and 32\% for the cc15, cc25, and snIa models, respectively, due to the reduction of the $^{84}$Sr($\gamma$,p)$^{83}$Rb rate.

The Monte Carlo variation performed in \cite{Rauscher} was repeated including the current experimental results. The MC variation factors were derived from the uncertainties as described in \cite{Rauscher}.
For the rates of the ${}^{83}$Rb(p,$\gamma$)${}^{84}$Sr reaction and its inverse, we adopted 0.06 and 1.94 for the lower limit and upper limit of the variation factor, respectively,
whereas 0.27 and 1.94 were used for the rate of the ${}^{84}$Kr(p,$\gamma$)${}^{85}$Rb reaction and its inverse.
It was found that the remaining, total uncertainty in the production factor of $^{84}$Sr is reduced to about half the previous value. The remaining uncertainty is shown in the form of 90\% CL error bars in Fig.\ \ref{fig:ccabuns}. It not only includes the cross section uncertainty from the
present measurement but stems from the combined uncertainties of all rates affecting the $^{84}$Sr abundance. No uncertainty is shown for the thermonuclear (SN Ia) supernova case because the previous nuclear uncertainty was already smaller than the size of the marker
in the figure.

It has been proposed that the elevated $^{84}$Sr abundances discovered in CAIs in 
the Allende meteorite \cite{Charlier} may be accounted 
for by $r$- and $s$-process variability in $^{88}$Sr production. While the increased production factors obtained in this work are not sufficient to reproduce these $^{84}$Sr abundances, increased production by a $\gamma$ process in explosions of massive stars and/or thermonuclear supernovae may ease the explanation of these abundances. To address this question in more detail, extensive Galactic chemical 
evolution models are required. This is beyond the scope of the current paper.

\section{\label{sec:concl}Conclusions}

In summary, we carried out the first direct measurement of the cross section of an astrophysical $\gamma$ process reaction in the Gamow window using a radioactive beam. A novel experimental method facilitated measurements of the partial cross section of the $^{83}$Rb(p,$\gamma)^{84}$Sr reaction at energies of $E_{\mathrm{cm}}$ = 2.260(7) and 2.386(23) MeV, indicating that the thermonuclear reaction rate is lower than that predicted by statistical model calculations. These predictions depend strongly on the proton width that, in turn, is determined by the penetration through the Coulomb barrier. Presently, it is not entirely clear how theory could exactly reproduce the central value of the measured data point at  $E_{\mathrm{cm}} = 2.386(23)$ MeV. Further investigations using data across a wider energy range within the Gamow window may help to better understand the differences.

With a smaller reaction cross section, the abundance of $^{84}$Sr produced during the astrophysical $\gamma$ process is larger than previously expected but still not large enough to explain the observation of elevated levels of $^{84}$Sr discovered in meteorites. Nevertheless, increased production in core-collapse and thermonuclear supernovae may impact Galactic chemical evolution models and change the requirements for additional sources of $^{84}$Sr.

Given the discrepancy between the present experimental measurements and theoretical predictions, we encourage the further study of $\gamma$-process reactions involving unstable projectiles. These reactions may hold the key to understanding the measured abundances of several $p$ nuclides from various sources in our Galaxy.% and, therefore, warrant experimental investigation.

\begin{acknowledgments}
The authors acknowledge the generous support of the Natural Sciences and Engineering Research Council of Canada. TRIUMF receives federal funding via a contribution agreement through the National Research Council of Canada.
The GRIFFIN infrastructure was funded jointly by the Canada Foundation for Innovation, the Ontario Ministry of Research and Innovation, the British Columbia Knowledge Development Fund, TRIUMF, and the University of Guelph. C.N. was supported by the U.S. Department of Energy under contract no. DE-FG02-93ER40789. UK personnel were supported by the Science and Technologies Facilities Council (STFC). T.R. acknowledges support by the European COST action ``ChETEC'' (CA16117). N.N. acknowledges support by JSPS KAKENHI (19H00693, 20H05648, 21H01087). A.P. acknowledges support from the State of Hesse within the Research Cluster ELEMENTS (Project ID 500/10.006).

\end{acknowledgments}

\bibliography{83Rb_pg}

\end{document}